# Electron-vibrational interaction in the 5d states of $Eu^{2+}$ ions in $Sr_{6-x}Eu_xBP_5O_{20}$ (x=0.01; 0.03; 0.05; 0.07; 0.09; 0.11; 0.13; 0.15)

Dejian Hou[a], C.-G. Ma[b], Hongbin Liang[a], M.G. Brik[c, *]

[a] *MOE Laboratory of Bioinorganic and Synthetic Chemistry, KLGHEI of Environment and Energy Chemistry, State Key Laboratory of Optoelectronic Materials and Technologies, School of Chemistry and Chemical Engineering, Sun Yat-sen University, Guangzhou 510275, People's Republic of China*

[b] *College of Mathematics and Physics, Chongqing University of Posts and Telecommunications, Chongqing 400065, People's Republic of China*

[c] *Institute of Physics, University of Tartu, Riia 142, Tartu 51014, Estonia*

**Abstract**

In the present paper we report on the combined experimental and theoretical study of the $Sr_{6-x}Eu_xBP_5O_{20}$ (x=0.01; 0.03; 0.05; 0.07; 0.09; 0.11; 0.13; 0.15) phosphors. Details of the samples preparation and spectroscopic measurements are followed by the analysis of the room-temperature absorption and emission spectra, which yielded the main parameters of the electron-phonon coupling, such as Huang-Rhys factor, Stokes shift, effective phonon energy, and zero-phonon line position were determined for the first time for the studied system. The obtained parameters were used to model the emission band shapes, which perfectly reproduce the experimental results for all samples.

---

[*] Corresponding author. Tel.: +372 7374751 E-mail: brik@fi.tartu.ee

# 1. Introduction

Luminescent materials firmly occupy a very important place in our everyday life. It is impossible to imagine the modern world without numerous devices, whose performance is based on the efficient emission produced by the controlled defects created within their volume. The rare earth (RE) ions are very often used as such defects to produce multicolored luminescence. Depending on the ion employed and electronic configurations involved into the absorption/emission transitions, a very wide spectral region can be covered, from infra-red to vacuum ultraviolet. If the intraconfigurational parity forbidden $4f^n$-$4f^n$ transitions are used, then the emission comes out as very sharp spectral lines. If the interconfigurational parity allowed $4f^n$-$4f^{n-1}5d$ transitions are excited, then the observed emission is characterized by very broad spectral bands, which allow for fine tuning and selection of the necessary emission wavelength. In addition, for some ions, like $Eu^{2+}$, the $4f^n$-$4f^{n-1}5d$ transitions are characterized by a very long lifetime, which is of paramount importance for production and successful applications of the persistent phosphors.

$Eu^{2+}$ ion has the $4f^7$ electron configurations in the ground state and the $4f^65d$ electron configuration in the excited state. The allowed 5d → 4f transitions from the $4f^65d^1$ configuration to the $4f^7$ configuration give rise to broad emission band, which strongly depends on the crystal strength surrounding of $Eu^{2+}$ ions. The reason for that is that the 5d electron shell is not screened like the 4f electron shell and, as such, can effectively interact with the nearest environment of the $Eu^{2+}$ ion incorporated into a host matrix. The 4f-5d transition bands of $Eu^{2+}$ can therefore be well controlled through such factors as the change of the crystal field, the site symmetry and the coordination environment [1]. The luminescence properties of the $Eu^{2+}$ doped inorganic phosphors have been extensively investigated from the perspective of their potential applications for lightings and displays [1-3]. For example, $(Ba, Sr)_3BP_3O_{12}$: $Eu^{2+}$ can serve as a promising green and bluish-white phosphor for near-UV excited LED; $Sr_8(Si_4O_{12})Cl_8$: $Eu^{2+}$ is a potential cyan-emitting phosphor in wide gamut 3D-PDP and 3D-FED; $Eu^{2+}$ doped $BaCa_2MgSi_2O_8$ phosphor may be suitable for tricolor lamps.

Another suitable material, $Sr_6BP_5O_{20}$, crystallizes in space group $I$-4$c$2, the $Sr^{2+}$ atoms occupy two sites (the so called Sr1 and Sr2), which are coordinated by nine (Sr1) and eight (Sr2) $O^{2-}$ atoms, respectively. In this crystal there are also five different oxygen sites. It is worthwhile noting that the Sr2 ions are surrounded by four O2 ions, by two O4 and two O5 ions, at the averaged Sr-O distance of 2.6955 Å. Environment of the Sr1 positions is much less symmetric, it consists of two O1 ions, one O2 ion, three O3 ions, two O4 ions and one O5 ion, with the averaged Sr-O distance of 2.7213 Å. The anionic layer consists of $BO_4$ and $PO_4$ tetrahedra [4]. Fig. 1 shows different surrounding of the strontium ions at both available crystallographic positions with indication of the Sr-O chemical bonds (in Å).

The luminescence properties and potential applications of $Eu^{2+}$ doped $Sr_6BP_5O_{20}$ for LED, PDP, and FED were reported [5-8]. The 4f-5d transition bands of $Eu^{2+}$ consist of zero-phonon lines and broad vibronic progressions, and thus are usually difficult to resolve and analyze experimentally.

In the present paper we continue studies of the $Sr_6BP_5O_{20}$:$Eu^{2+}$ phosphor. Starting from the experimental absorption and emission spectra of $Sr_6BP_5O_{20}$:$Eu^{2+}$, we have estimated the Huang-Rhys factor, the energy of an effective phonon interacting with the $Eu^{2+}$ 5d states, and the zero-phonon line (ZPL) position. These parameters for the title system are reported for the first time, to the best of our knowledge. As a check of reasonability of the obtained parameters, the $Eu^{2+}$ 5d-4f emission band was modeled and compared with the corresponding experimental result; good agreement was demonstrated, which justifies validity of the estimated and reported EVI parameters.

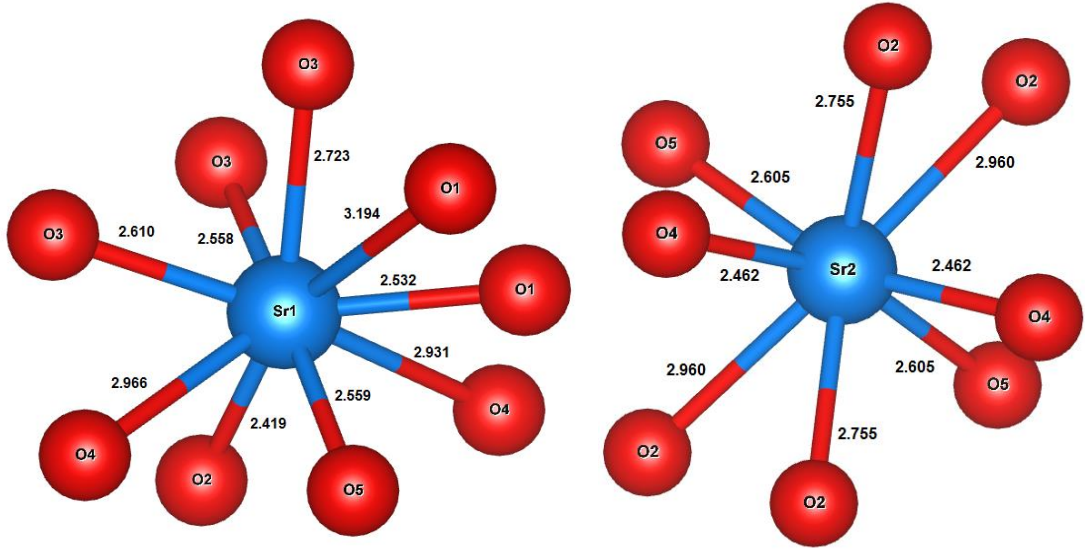

Fig. 1. Sr1 and Sr2 positions in $Sr_6BP_5O_{20}$. The oxygen ions are labeled according to their crystallographic positions. The Sr-O distances (in Å) are also given.

## 2. Theoretical background

The interconfigurational 5d-4f transitions between the energetic states of the $4f^{n-1}5d$ and $4f^n$ electron configurations manifest themselves as broad bands in the emission and absorption/excitation spectra of materials containing rare earth ions. These transitions are dipole allowed, therefore, their intensity is much higher than that one of the intraconfigurational 4f-4f transitions. Since the 4f-5d transitions involve the outer 5d shells of RE ions (either as an initial, or a final state), which interact strongly with the vibrating environment in the host's crystal lattice, the problem of a detailed analysis of the electron-vibrational interaction (EVI) between the 5d states and crystal lattice vibrations and its influence on the overall appearance of the optical spectra is an important task.

There are three main parameters, which describe the EVI in the impurity centers. They are as follows: the Stokes shift $\Delta E_S$ (the energy difference between the lowest in energy absorption and emission peaks corresponding to the same electronic states), the Huang-Rhys factor $S$ (which is proportional to $\Delta E_S$) and the effective phonon energy $\hbar\omega$, mainly responsible for the EVI. All these parameters arise from the shift of the equilibrium positions of the potential energy surfaces in the ground and excited states, as shown in the framework of a simple one configurational coordinate model [9]. The values of $S$ and $\hbar\omega$ can be estimated using the following equations:

$$\Delta E_S = (2S-1)\hbar\omega \qquad (1)$$

$$\Gamma(T) = \sqrt{8\ln 2}\,\hbar\omega \left[ S \coth\left(\frac{\hbar\omega}{2kT}\right) \right]^{1/2} \qquad (2)$$

where the last expression describes the full width $\Gamma(T)$ at half maximum (FWHM) of the emission band as the function of the absolute temperature $T$. The easiest way to find the values of $S$ and $\hbar\omega$ from the system of non-linear equations (1) - (2) is to solve these equations graphically with the values of $\Gamma(T)$ and $\Delta E_S$ extracted from the experimental emission and absorption/excitation spectra.

As an additional check of reasonability of the obtained in this way values of $S$ and $\hbar\omega$, the emission band shape can be modeled and compared with the experimental emission spectrum. The intensity $I$ of the emission band at energy $E$ can be approximated by the following expression [9]:

$$I = \frac{e^{-S} S^p}{p!}\left(1 + S^2 \frac{e^{-\hbar\omega/kT}}{p+1}\right), \quad p = \frac{E_0 - E}{\hbar\omega}, \qquad (3)$$

where $E_0$ is the zero phonon line (ZPL) energy, which is located closely to the point of intersection of the excitation and emission spectra, and $p$ is the number of the effective phonons involved into the emission transition. All other quantities entering Eq. (3) have been described above. In the ideal case, when the emission and absorption spectra are the mirror images of each other, the ZPL position is determined exactly and unambiguously by the point of intersection of both spectra. However, for a vast majority of the real optical systems the above-mentioned mirror symmetry of the emission/absorption spectra is broken, and experimental determination of the ZPL position is a rather complicated problem, which would imply getting high-resolved low-temperature spectra as a first and necessary requirement.

Application of Eqs. (1) - (3) allows for estimation of the EVI parameters, including the ZPL position. This model, even being quite simple, was proven to be successful in description of the EVI in the 5d states of $Eu^{2+}$ ion in a number of hosts [10-12]

## 3. Experimental details

The $Eu^{2+}$ ions doped samples $Sr_{6-x}Eu_xBP_5O_{20}$ (x = 0.01, 0.03, 0.05, 0.07, 0.09, 0.11, 0.13, 0.15) were synthesized through a traditional high temperature solid state reaction technique. The starting materials are $SrCO_3$ (analytical reagent A. R.), $(NH_4)_2HPO_4$ (A. R.), $H_3BO_3$ (A. R.) and $Eu_2O_3$ (99.99 %). $H_3BO_3$ should be excess 60 mol% to compensate evaporation, and then we can get an XRD pure phase under this condition. After mixing and thoroughly grinding, the stoichiometric mixtures were sintered at 1020 °C for 2 hours in thermal-carbon reducing atmosphere.

The phase purity of the samples $Sr_{6-x}Eu_xBP_5O_{20}$ was measured by powder X-ray diffraction (XRD) with CuKα radiation on a D8 ADVANCE X-ray Diffractometer operating at 40 kV and 40 mA. The photoluminescence spectra were measured by FLS 920 equipped with a 450 W Xe lamp, TM300 excitation monochromator and double TM300 emission monochromators. All the measurements were carried out at room temperature.

## 4. Analysis of the excitation and emission spectra of $Eu^{2+}$ ions in $Sr_6BP_5O_{20}$:$Eu^{2+}$ and estimation of the EVI parameters

To check the phase purity of the samples, the powder X-ray diffraction (XRD) measurements for all samples were carried out. The results show that all samples are of single pure phase, and their powder XRD data are in agreement with the calculated standard card [8].

The room temperature excitation and emission spectra of $Eu^{2+}$ ions in $Sr_6BP_5O_{20}$ powders are shown in Fig. 2. The main spectral features can be unambiguously attributed to the $4f^7$-$4f^65d$ absorption and $4f^65d$-$4f^7$ emission transition of divalent europium ions, which substitute for the strontium ions.

The absorption spectra are characterized by a very broad intense band, which stretches from about 22000 cm$^{-1}$ to about 45000 cm$^{-1}$. Such a broad width of the band is due to the crystal field splitting of the 5d states of Eu; the splitting is very large and may suggest a very low symmetry of the local positions occupied by the $Eu^{2+}$ ions. The lowest 5d level was determined from the position of the first feature in the absorption spectra at about 25200 cm$^{-1}$ (denoted by an arrow in all spectra).

The emission spectra represent a wide structureless band with a maximum at about 20600-20800 cm$^{-1}$ (slightly depending on the dopant concentration) and FWHM of about 3200-3600 cm$^{-1}$.

In principle, the absorption spectrum could be decomposed into at least several individual Gaussian bands, to mimic the splitting of the Eu 5d states by crystal field. However, such decomposition is not unique, since it is practically impossible in this case to identify the ZPL lines corresponding to the transitions from the ground state to the higher located components of the Eu 5d states, which can overlap with vibronic transitions etc. This is the reason why we focused our attention on the lowest absorption peak only, whose position can be identified easily in each of the shown spectra.

The absorption and emission spectra shown in Fig. 2 provide all necessary data for solution of a system of Eqs. (1)-(2). Table 1 below contains the summary of the experimental data needed for the estimation of the EVI parameters, and the obtained values of the Stokes shift, Huang-Rhys factor, effective phonon energy, and ZPL position.

Table 1. The main spectroscopic and EVI parameters for the $Sr_{6-x}Eu_xBP_5O_{20}$ (x=0.01; 0.03; 0.05; 0.07; 0.09; 0.11; 0.13; 0.15) samples.

| | $1^{st}$ absorption max., cm$^{-1}$ | Emission max., cm$^{-1}$ | $\Delta E_S$, cm$^{-1}$ | FWHM, cm$^{-1}$ | $S$ | $\hbar\omega$, cm$^{-1}$ | ZPL, cm$^{-1}$ Fit, Eq. (3) | ZPL, cm$^{-1}$ Exp. |
|---|---|---|---|---|---|---|---|---|
| x=0.01 | 25200 | 20833 | 4367 | 3188 | 3.78 | 676 | 22850 | 23165 |
| x=0.03 | 25242 | 20873 | 4369 | 3224 | 3.65 | 693 | 22900 | 23130 |
| x=0.05 | 25240 | 20800 | 4400 | 3109 | 3.65 | 693 | 22900 | 23060 |
| x=0.07 | 25242 | 20765 | 4477 | 3152 | 3.98 | 632 | 22900 | 23090 |
| x=0.09 | 25200 | 20900 | 4300 | 3108 | 3.81 | 648 | 22900 | 23130 |
| x=0.11 | 25243 | 20875 | 4368 | 3181 | 3.84 | 654 | 22900 | 23095 |
| x=0.13 | 25200 | 20900 | 4300 | 3551 | 3.00 | 851 | 22950 | 23130 |
| x=0.15 | 25279 | 20945 | 4334 | 3699 | 2.96 | 874 | 22950 | 23060 |

One can note that for the highly doped samples the effective phonon energy is increased. This is due to increase of the FWHM (the Stokes shift values are nearly the same for all samples). We attribute such FWHM increase to considerable changes of the host's phonon spectrum at high concentrations of europium and appearance of some local modes.

With the values of $S$ and $\hbar\omega$ from Table 1, we succeeded in modeling the emission band shape using Eq. (3). Only one parameter - namely, the ZPL position $E_0$, was allowed to vary freely until the best agreement between the experimental and calculated band shapes is reached. In this way, the ZPL position for the 5d - 4f emission of $Eu^{2+}$ in $Sr_6BP_5O_{20}$ was estimated to be at about 22900 cm$^{-1}$, being only slightly dependent on the europium concentration. For all eight samples considered, the shapes of the experimental and simulated emission bands practically coincide, which justifies the values of *S* and $\hbar\omega$ obtained from Eqs. (1) - (2).

As it was mentioned above, the ZPL position can be also estimated as the point of intersection of the excitation and emission spectra. In this case one neglects certain deviation from the mirror symmetry of the absorption and emission spectra. Even if the emission and absorption spectra in Fig.2 are not the mirror images of each other, the position of the point of their intersection (given in Table 1) is reasonably close to the corresponding ZPL positions obtained from the emission band shape modeling.

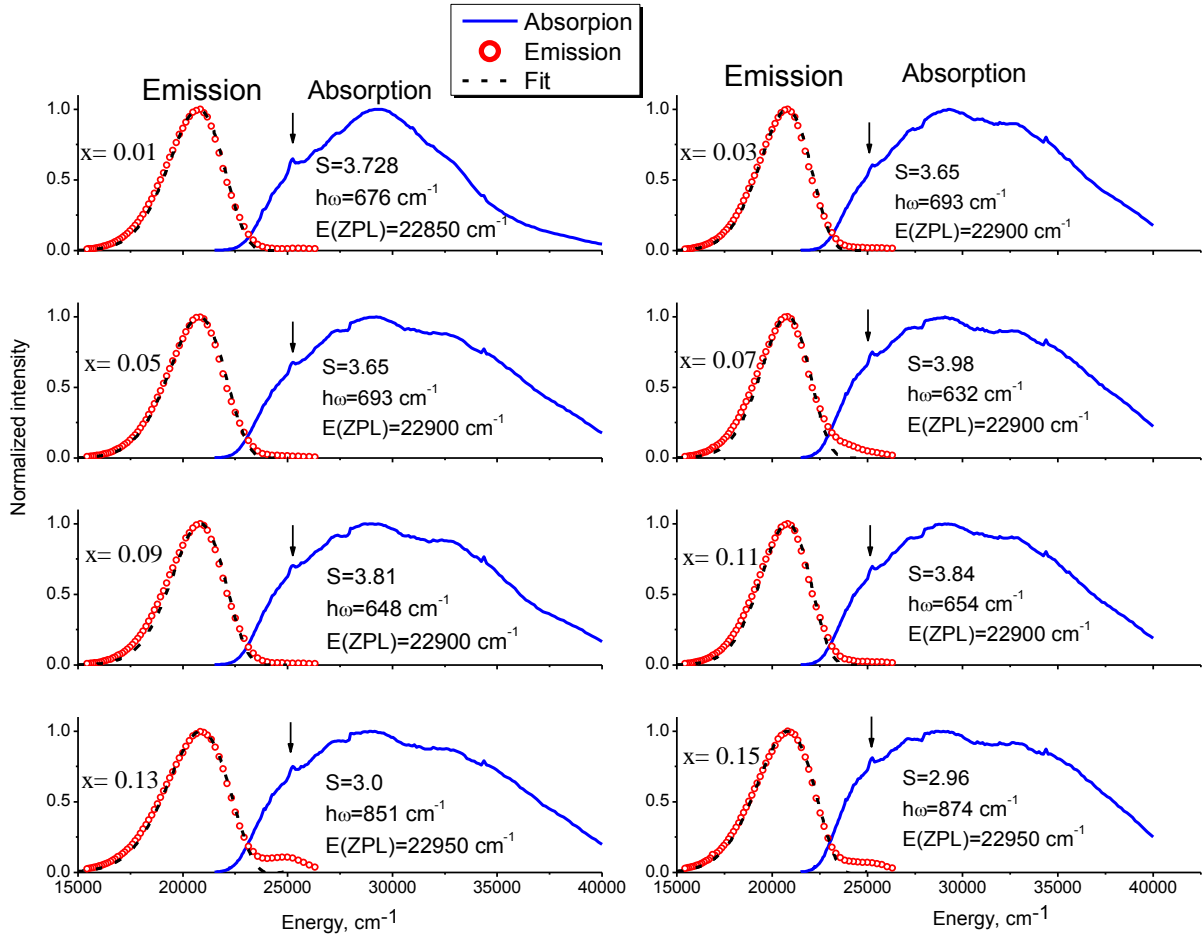

Fig. 2. Room temperature absorption and emission spectra of $Sr_{6-x}Eu_xBP_5O_{20}$ in comparison with the calculated emission band shape. The Eu content x is given in each plot.

## 5. Conclusions

Details of preparation and spectroscopic studies of the $Sr_{6-x}Eu_xBP_5O_{20}$ phosphor (with different concentration of the Eu dopants from x=0.01 to x=0.15) are reported in the present paper. The structure of the prepared samples was confirmed by the XRD measurements to ensure that no other phase exists in the samples. The room temperature absorption and emission spectra of the prepared phosphors were studied using the single configurational coordinate model. In this way, the main parameters of the electron-vibrational interaction, such as the Stokes shift, Huang-Rhys factor, effective phonon energy, and zero-phonon line position, were derived. They effectively describe interaction of the $Eu^{2+}$ 5d states with the crystal lattice environment. As a proof of reasonability of the obtained parameters, the emission band shape was modeled for each sample to yield very good agreement with the experimental spectra, both from the point of view of the emission band width and position of its maximum.

**Acknowledgement:**
Hongbin Liang is financially supported by National Natural Science Foundation of China (Nos. 10979027, 21171176 and U1232108), and Guangzhou Municipal Science and Technology Project (2013Y200118). C.-G. Ma acknowledges the financial support by National Natural Science Foundation of China under Grant No 11204393. M.G. Brik acknowledges the support from the European Social Fund's Doctoral Studies and Internationalization Programme DoRa and the European Regional Development Fund (Centre of Excellence 'Mesosystems: Theory and Applications', TK114).